\documentclass[twocolumn,pra]{revtex4-1}
\usepackage[active]{srcltx}
\usepackage{graphicx}
\usepackage{bm}
\usepackage{textcomp}
\usepackage{units}
\usepackage{amsmath}
\usepackage{tikz}

\begin{document}
 
\bibliographystyle{unsrt}
\def\fr#1#2{\frac{\textstyle #1}{\textstyle #2}}
\def\rd{{\rm d}}
\def\re{{\rm e}}
\def\ri{{\rm i}}
\def\iO{{\it \Omega}}
\title{A proposal to classify the radian as a base unit in the SI}

\author{Peter J. Mohr}
\email{mohr@nist.gov}

\author{William D. Phillips}
\email{wphillips@nist.gov}

\affiliation{National Institute of Standards and Technology,
Gaithersburg, MD 20899, USA}


\begin{abstract} 

We propose that the SI be modified so that the radian is a base unit.
Some of the details and consequences of such a modification are
examined.

\end{abstract}

\maketitle

\date{\today}

\section{Introduction}

\vspace{-0.3 cm}

\subsection{Background}

\vspace{-0.3 cm}

When the International System of Units (SI) was established by
Resolution 12 of the General Conference on Weights and Measures (CGPM)
in 1960, units were classified into three classes, base units,
supplementary units, and derived units.  The category of supplementary
units consisted of the radian for plane angle and the steradian for
solid angle.

In 1995, Resolution 8 of the 20$^{\rm th}$ CGPM stated the decision ``to
interpret the supplementary units in the SI, namely the radian and the
steradian, as dimensionless derived units, the names and symbols of
which may, but need not, be used in expressions for other SI derived
units, as convenient.''

\vspace{-0.4 cm}

\subsection{Current status of plane angle and solid angle}

\vspace{-0.3 cm}

As a result of the reclassification of the radian and steradian as
dimensionless derived units, they are listed in the SI Brochure
\cite{2006SI} in Table 3, ``Coherent derived units in the SI with
special names and symbols,'' as the unit for plane angle expressed as
m/m in terms of SI base units and as the unit for solid angle expressed
as m$^2$/m$^2$ in terms of SI base units, respectively.  In a footnote
to Table 3, it is stated that: ``{\em The radian and steradian are
special names for the number one} that may be used to convey information
about the quantity concerned. In practice the symbols rad and sr are
used where appropriate, but the symbol for the derived unit one is
generally omitted in specifying the values of dimensionless
quantities.''

In the above quotation, the italics are added for emphasis; we view the
italicized statement as nonsense.  Furthermore, this practice has led to
errors in published results for physical quantities involving angles and
frequencies.

\vspace{-0.4 cm}

\subsection{Proposal}

\vspace{-0.3 cm}

It is proposed here that angles be considered to have dimension (the
dimension of angle) and that therefore the radian no longer be
considered a dimensionless derived unit and instead be reclassified as a
base unit.  It should be considered the coherent SI unit for plane angle
and for phase, sometimes called phase angle \footnote{Here the term
phase refers to a physical phase as in quantum mechanics or
electrodynamics.  In mathematics, particularly complex variable theory,
phase is regarded as a dimensionless number as is length.}.  This has
implications for the units of frequency, as explained below.  

The steradian is then a derived unit with the unit rad$^2$.  

\section{supplemental material}

\subsection{Plane angle and phase as physical quantities}

Plane angles and phase angles have properties similar to other
measurable physical quantities.  In particular, the value of a plane
angle or phase angle $\theta$ can be written in the SI as 
\begin{eqnarray}
\theta = \{\theta\}[\theta]
\label{eq:angle}
\end{eqnarray}
where $\{\theta\}$ is the {\it numerical value} of the angle in the unit
radian and $[\theta]$ is the unit rad.  For a plane angle, the numerical
value of the angle between two intersecting straight lines is the ratio
of the length of the arc $s$ between the lines of a circle centered at
the vertex of the angle to the radius $r$ of the circle $\{\theta\} =
s/r$ and the unit is $[\theta] =$ rad.  In the current SI, the unit
$[\theta]$ may be either omitted or replaced by the number $1$.

It is sometimes said that angles are dimensionless quantities, because
(as suggested in the SI Brochure) the ratio $s/r$ is a length divided a
length and is therefore just a number.  In fact, the number in curly
brackets for {\it any} physical quantity is just a number, but that does
not make the physical quantity itself dimensionless.

For example, consider a two-meter long table.  The numerical value of
the length of the table in meters is the ratio of the length of the
table to the length of a meter stick.  This does not mean that the
length of the table is a dimensionless quantity; it has the dimension of
length with the unit of meter.  In the same way for a two-radian angle,
the numerical value of the angle in radians is the ratio of the length
of the subtended arc to the radius.  Similarly, this does not mean that
angle is a dimensionless quantity; it has the dimension of angle with
the unit of radian.  

It is an essential tenet of this note that the radian should not be
considered a dimensionless unit in the SI; it should have dimension
angle in the SI; it is not derived from other base units, and therefore
should be a base unit itself, rather than a derived unit as in the
present SI.

\subsection{Relations to non-SI units for angles}

As with any physical quantity, angles may be expressed in units other
than coherent SI units.  However, units other than the radian cannot
also be coherent SI units if the radian is.  Some other units for angles
are degrees, gradians, turns, cycles, and revolutions.  Angles range
anywhere from zero to a complete revolution; angles greater than $\pi$
rad are sometimes called reflex angles.  A complete revolution is an
angle of $2\pi$ rad.  

Some relations to the non-SI units are
\begin{eqnarray}
180^\circ = \pi \mbox{ rad}
\\
100 \mbox{ grad} = \frac{\pi}{2} \mbox{ rad}
\\
1 \mbox{ turn} = 2\pi \mbox{ rad}
\\
1 \mbox{ cycle} = 2\pi \mbox{ rad}
\label{eq:crad}
\\
1 \mbox{ revolution} = 2\pi \mbox{ rad}
\end{eqnarray}

\subsection{Periodic phenomena}

A physical phenomenon that repeats in time over a regular time interval
is periodic or cyclic.  One repetition of the phenomenon that happens in
a period is a cycle.  The rate of repetitions is a physical quantity
called the frequency.  The numerical value of the frequency will depend
on the units in which it is expressed.

We recommend that the SI states that the value of any frequency of
a periodic phenomenon include the complete unit such as rad/sec or Hz,
but never just s$^{-1}$.

An example is a turning bicycle wheel.  If the wheel rotates by 2 radians in
one second, its rotational frequency in SI units is 2 rad/s.

Another example is electromagnetic radiation, where the electric field
vector undergoes a periodic repetition of one million cycles per second.
This frequency is
\begin{eqnarray}
\nu = 1 \mbox{ MHz}, 
\label{eq:nu}
\end{eqnarray}
where MHz is one million cycles per
second.  Since the electric field vector may be described as
\begin{eqnarray}
\bm E \sin{\omega t},
\label{eq:sine}
\end{eqnarray}
where the time dependence of the vector repeats when $\omega t$
increases by $2\pi$.  (Here we take $\omega t$ to represent the
numerical value of the product.)  If $t$ is expressed in seconds, then
$\omega$  must be expressed in the coherent unit rad/s:
\begin{eqnarray}
\omega=2\pi\times 10^6 \mbox{ rad/s}.
\label{eq:omega}
\end{eqnarray}

\subsection{Difference with the current SI}

There are two principal differences between the formulation above and
the current SI.  One consequence of writing plane angles and phase
angles as quantities with explicit units is that an ambiguity present in
the current SI is prevented.  In particular, in the current SI, since
the SI unit radian is considered a dimensionless unit, it may be
omitted.  The same is true for the non-SI unit "cycle," which is omitted
when the currently permitted replacement Hz $\rightarrow$ s$^{-1}$ is
made.  As a result of these prescriptions, the units on both sides of
Eq.~(\ref{eq:crad}) may be omitted leading to an inconsistent result.

The other principal difference that results from regarding angles as
quantities with units that may not be dropped is that Hz, {\it i.e.}
cycles/second, explicitly includes the non-SI unit cycle and is
therefore not a coherent SI unit.  Instead, radians/second is the
coherent SI unit for frequency.  As a consequence, Hz should be
considered a unit ``permitted for use with the SI.''

\subsection{Mathematics vs physics}

In texts and monographs on mathematics, particularly calculus and
complex variable theory, there is little or no discussion of units.
Lengths and angles are simply numbers with no units and with no mention
of meters or radians. As a result, mathematical functions such as
trigonometric functions, the exponential function, spherical Bessel
functions, spherical harmonics, or any number of other mathematical
functions are defined as functions of a dimensionless real or complex
number.

On the other hand, in physics, it is necessary to include units in order
to make contact with measurements. The values of physical quantities are
given by a number times a unit. Thus when, for example, the sine
function is used in physics, the argument often is written as a quantity
whose units are understood to be radians, as in Eq.~(\ref{eq:sine}).
However, since the sine function can be defined by its Taylor series,
and the terms must be homogeneous in their units, the argument is
necessarily a number with no unit.

This leads to a problem with the use of the radian as a unit in the
argument of functions that are defined as having dimensionless
arguments. An unambiguous resolution of this conflict would be to write
the argument of mathematical functions as the numerical value of the
angle.  That is, to write the sine function in Eq.~(\ref{eq:sine}) as
sin $\{\omega t\}$, where $\omega t = \{\omega t\}$ rad.

However, in scientiﬁc publications, functions are written with arguments
without the curly brackets, and it would be too disruptive to the common
practice to insist that curly brackets should be present. So a
compromise would be to recommend that when angles are present in
mathematical functions and expressed in units of radians, the curly
brackets be omitted.  By the same token, it should also be recognized
that when quantities that are the arguments of functions such as the
sine function are used in equations outside of those functions, the
radian unit must be explicitly restored.

\subsection{Illustrative example}

A simple example that illustrates the points made above is the physics
of a harmonic oscillator. The equation for a harmonic oscillator is
\begin{eqnarray}
m \, \frac{{\rm d}^2}{{\rm d}t^2}\, x(t) + kx(t) = 0,
\end{eqnarray}
where $m$ is the mass and $x(t)$ is the coordinate of the body and $k$
is the spring constant of the spring. The solution for the motion of the
body is
\begin{eqnarray}
x(t) = x_0 \sin\left\{\sqrt{\frac{k}{m}} \ t + \phi \right\} ,
\end{eqnarray}
where $x_0$ is the amplitude of the motion and $\phi$ is a phase that
depends on the boundary condition. The curly brackets emphasize that the
argument of the sine function is the numerical value of the enclosed
physical quantities.  The frequency of the oscillations is
\begin{eqnarray}
\mbox{frequency} = \sqrt{\frac{k}{m}} =
\left\{\sqrt{\frac{k}{m}}\right\} \mbox{s}^{-1} .
\end{eqnarray}
In the current SI, where both rad/s and Hz may be expressed as s$^{-1}$,
this could be taken to mean both
\begin{eqnarray}
\mbox{frequency} = 
\left\{\sqrt{\frac{k}{m}}\right\} \mbox{Hz}
\qquad\mbox{(incorrect)}
\end{eqnarray}
and
\begin{eqnarray}
\mbox{frequency} = 
\left\{\sqrt{\frac{k}{m}}\right\} \mbox{rad s}^{-1}
\qquad\mbox{(correct)}.
\label{eq:corr}
\end{eqnarray}
The choice of Eq.~(\ref{eq:corr}) as the correct expression of the
frequency is evident from the rule (see above) that when the argument of
a trigonometric function (or the imaginary argument of an exponential
function) is used outside of such functions, the unit radian must be
explicitly restored.

\subsection{On the nature of units}

In this brief note, we have proposed that the radian be viewed as a base
unit in the SI.  In this section, we clarify some issues relating to
units.

Values of most physical quantities $q$ can be written as
\begin{eqnarray}
q = \{q\}[q],
\end{eqnarray}
where $\{q\}$ is the numerical value of the quantity expressed in the
unit $[q]$.  As seen in Eq.~(\ref{eq:angle}), angles fall into this
category.  Another category of physical quantity, examples of which are
the proton-electron mass ratio and the fine-structure constant, are
indeed dimensionless and unitless.  They are simply numbers.  Angles do not
fall into this category.

Concerning units for frequency, some of the ambiguity and confusion over
frequency arises from the idea that frequency may be classified into
types of frequency with different units being used for different types.
The prime example is Hz vs rad/s which might be considered different
types of frequency and so are expressed in different units.  However,
frequency is a general concept and classifying types of frequency leads
to confusion.  For frequency, there is one coherent unit, which is
rad/s, and which applies to all types of frequency whether cycles of
electromagnetic radiation or rotations of a wheel or heartbeats.  In the
latter case, the role of radians per second is easily seen by
considering a Fourier decomposition of the pattern of the heartbeats and
realizing that the fundamental frequency can be viewed as the rate of
change of the phase of a sine function with the argument being the
numerical value of the phase angle expressed in rad.

This is analogous to energy, for example.  There are different types of
energy, mechanical energy, energy of electromagnetic radiation,
gravitational potential energy, but these are recognized to be various
forms of energy, and so they all have the same unit, joule.  Of course,
this is based on physical principles and may be counter-intuitive to
people unfamiliar with these principles, but nevertheless is well
understood to be true.

While in our proposal the radian becomes a base unit and angles have
dimension, we recognize that historic and common practice often drop the
unit radian where it would otherwise appear.  With the exception of
frequencies and only if no confusion can result, we would continue to
allow the unit radian to be dropped.

In the new SI, expected to be adopted in 2018, all units are defined by
giving an exact value to a fundamental physical constant, although not
all the definitions strictly adhere to that designation.  In the case of
angle, one could specify the unit radian by fixing the numerical value
of a right angle in radians by writing
\begin{eqnarray}
\theta_\perp
 = \frac{\pi}{2} \mbox{ rad},
\end{eqnarray}
where $\theta_\perp$ a right angle.

This note focuses on angles, radians and related quantities, because the
ambiguity in the SI associated with these concepts has often led to
errors and confusion.  Other similar quantities, such as counting units,
may need to be addressed in a similar manner, but this is beyond the
scope of this note.

Finally, we note that in the new SI, the distinction between base units
and derived units may be unnecessary.  The same may be true of the
distinction between dimensions and units.


\end{document}